# Into the Unknown: From Structure to Disorder in Protein Function Prediction

Đesika Kolarić[1,2], Chi Fung Willis Chow[3,4], Rita Zi Zhu[3,4],
Agnes Toth-Petroczy[3,4,5]*, T. Reid Alderson[1,2]*, Iva Pritišanac[6]*

1. Helmholtz Munich, Institute of Structural Biology, Molecular Targets and Therapeutics Center, 85764 Neuherberg, Germany
2. Technical University of Munich, TUM School of Natural Sciences, Department of Bioscience, Bavarian NMR Center, 85747 Garching, Germany
3. Max Planck Institute of Molecular Cell Biology and Genetics, Pfotenhauerstrasse 108, 01307 Dresden, Germany.
4. Center for Systems Biology Dresden, Pfotenhauerstrasse 108, 01307 Dresden, Germany.
5. Cluster of Excellence Physics of Life, TU Dresden, 01062 Dresden, Germany.
6. Medical University of Graz, Division of Medicinal Chemistry, Otto Loewi Research Center, 8010 Graz, Austria

*correspondence to: toth-petroczy@mpi-cbg.de, reid.alderson@helmholtz-munich.de, iva.pritisanac@medunigraz.at

## Abstract

Intrinsically disordered regions (IDRs) account for one-third of the human proteome and play essential biological roles. However, predicting the functions of IDRs remains a major challenge due to their lack of stable structures, generally faster sequence evolution relative to folded domains, and context-dependent behavior. Many predictors of protein function neglect or underperform on IDRs. Recent methodological advances, including protein language models, alignment-free approaches, and IDR-specific methods, have revealed conserved bulk features and motifs within IDRs that are linked to function. This review highlights emerging computational methods that chart the sequence-function relationship in IDRs, outlines critical challenges in IDR function annotation, and proposes a community-driven framework to accelerate interpretable functional predictions for IDRs.

**A Whole New World: The functional significance of IDRs**

Proteins orchestrate and drive nearly every aspect of biology. The traditional structure-function paradigm posits that a stable protein structure is a prerequisite for function. However, over 30% of eukaryotic proteomes consist of intrinsically disordered regions (IDRs)[1], which challenge the structure-function paradigm and populate ensembles of dynamic, interconverting conformations[2, 3]. In humans, more than 60% of proteins contain an IDR, and thus most proteins consist of a mixture of order and disorder, while fully disordered proteins, or IDPs, comprise about 5% of human proteins[4]. Fundamentally, the interplay between order and disorder, juxtaposing a rigid and stable fold with conformationally dynamic regions, has been linked to the evolvability, robustness, and innovativeness of proteins[5–7]. The link between disorder and the emergence of new protein function is evident from the enrichment of IDRs in 'young' *de novo* genes[8] and unannotated coding sequences[9].

IDRs and IDPs (henceforth, IDRs) play essential roles throughout cellular biology, from regulating RNA transcription and protein translation to regulating transmembrane channels and serving as hubs in protein-protein interaction networks[10, 11]. The functional significance of IDRs is highlighted by their enrichment in genes linked to various diseases, including cancer, neurodegenerative disorders, and type-2 diabetes[12, 13]. While 10-20% of pathogenic variants fall within IDRs, the top variant effect predictors including recent deep learning algorithms have lower sensitivity in identifying pathogenic variants in disordered regions, and overpredict them as benign variants[14]. IDRs are also frequently associated with alternatively spliced isoforms[15] and post-translational modifications (PTMs)[16], thereby modulating functional diversity and creating a multitude of proteoforms. At the molecular level, IDRs engage in dynamic interactions that are often transient but specific, leveraging both short linear motifs (SLiMs)[17] and compositionally biased regions to impart specificity. Protein-binding residues within IDRs, known as molecular recognition features (MoRFs)[18], often adopt structural elements (typically helices) which facilitate binding[19]. Furthermore, IDRs frequently contribute multivalent interactions that promote liquid-liquid phase separation and the formation of biomolecular condensates[20]. The affinity and specificity of these IDR-mediated interactions are frequently fine-tuned by PTMs, enabling precise spatiotemporal regulation of protein function and localization[21].

However, the IDR sequence-to-function landscape remains poorly understood[22]. This is because IDR sequences are not constrained by the prerequisite to fold, which enables relatively rapid evolution when compared to their adjacent folded domains[23]. IDRs generally display more limited positional sequence conservation, which hinders functional inference from multiple sequence alignments (MSAs). While SLiMs tend to be positionally conserved, they typically comprise only a small fraction of intrinsically disordered sequences[24]. In addition, often the density of SLiMs rather than their exact position is the conserved feature[25]. Recent studies of IDRs have uncovered non-positional evolutionary conservation of cryptic motifs[26] and bulk sequence features, including amino-acid composition, charge distribution, and repeats[23, 27–30]. These bulk features have been directly linked to key biological functions, including a broad range of gene-ontology (GO) terms[23, 28, 30], as well as to molecular recognition[31–33] and phase separation[34][35][36] (**Box 1**).

In this review, we discuss recent methodological advances in deducing sequence-function relationships for IDRs, highlighting trends in machine learning and statistical modeling (**Figure 1**).

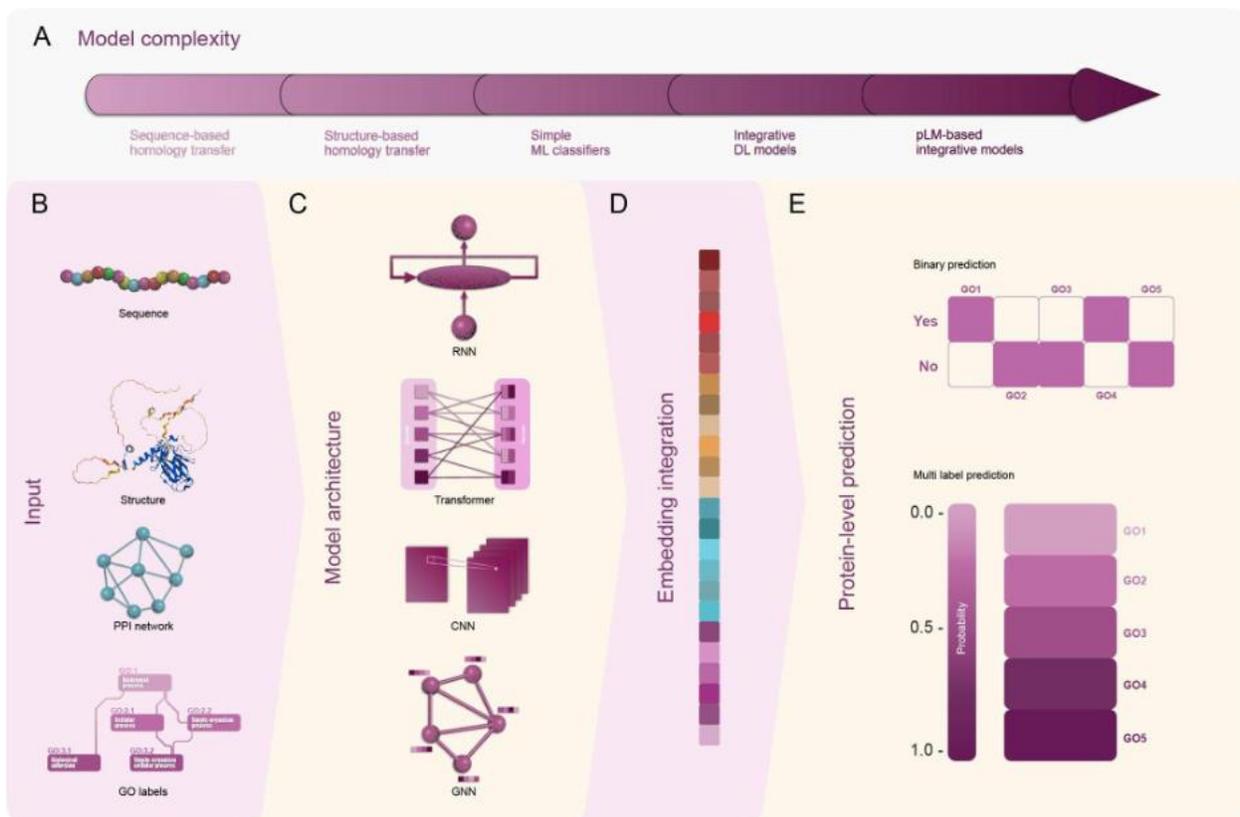

**Figure 1: Overview of advancements and current pipelines in protein function prediction.** (**A**) Summary of key advancements in protein function prediction, from early sequence-based annotation transfer methods to the emergence of protein language models. (**B-D**) General pipeline of current protein function prediction models. Various input data types (**B**) can be used individually or in combination within integrated models. Most models encode each input separately using different model architectures (**C**). The resulting embeddings are then combined in the final layer (**D**) for protein-level function prediction. (**E**) Model outputs can be either binary (yes/no, 1/0) for each Gene Ontology (GO) term or multi-label predictions, where each function is assigned a probability score indicating the likelihood of the protein performing that function.

**Under the Sea: Navigating the versatile functions of protein disorder**

An ordered or structured region typically exhibits a one-to-one sequence-to-function correspondence. This is because ordered domains autonomously fold into stable three-dimensional structures that have evolved to perform specific functions. Such structures tend to be evolutionary conserved and are defined as domains in structure- (CATH[37]/TED[38], SCOP[39], ECOD[40]) or sequence-based categorizations (Pfam[41], InterPro[42]). Example functions may include binding to an interaction partner to facilitate the assembly of a macromolecular complex, or catalyzing an enzymatic reaction.

Because of the structural dynamics and plasticity of IDRs, their sequence-to-function mapping often exhibits a one-to-many correspondence[43]. In addition, IDR functions can manifest at different levels, both globally and locally in the sequence **(Box 1)**. Local sequence features, for instance SLiMs, are often associated with specific molecular interactions. In addition to molecular recognition facilitated by SLiMs, IDRs can engage in distributed multivalent interactions that are driven by global sequence properties (e.g. charge distribution). While motifs are considered in specific stoichiometric interactions, distributed multivalent binding is mainly associated with biomolecular condensates[44]. Global sequence properties shape IDR conformation (flexibility, entropy, compactness) and biochemical properties (charge patterns, hydration shell), which also influence function and can be subject to selection[27, 45]. However, sequence alignments often fail to detect similarities in global properties across proteomes, as substantially different amino-acid sequences can achieve the same global bulk properties (*vide infra*). Finally, PTMs can influence both local and global properties and act as functional switches[46]. The intricate interplay of local and global sequence properties, as well as overlapping functional contributions within IDRs, complicates the prediction of IDR functions.

**Let It Go: Breaking free of alignments**

When compared to structured domains, sequence alignments of IDRs on average exhibit lower positional conservation and higher frequencies of insertions and deletions[24, 47], reflecting the relaxed or absent requirement for folding. Since sequence alignments primarily capture constraints that reflect inter-residue interactions that contribute to structure, MSAs are inherently less informative for IDRs[48]. However, we and others recently demonstrated that conditional structure in IDRs, acquired under specific conditions (e.g. binding), can be inferred from MSAs[49] and by AlphaFold2 for a subset of IDRs (approximated to about 15% of human IDRs)[50][51]. This suggests that numerous IDRs could function through other mechanisms that do not require a conditional fold. The inability of alignments to capture conservation of global sequence features (**Box 1**) necessitates alternative approaches that would account for the flexible nature of IDRs while identifying functionally relevant sequence constraints in order to fully extract the information encoded in IDR sequences.

Protein language models (PLMs) provide alternatives to alignment-based methods for IDRs (**Figure 1**), since they do not rely on homology or structural information to learn evolutionary and structural context at the residue, single-sequence and multiple-sequence levels. However, existing PLMs and PLM-based models are rarely trained or evaluated on IDRs. Some recent models have begun addressing this gap by training PLMs on large, diverse sequence datasets and subsequently fine-tuning them for IDR-related tasks, which has enabled predictions of disorder[52] and some biophysical properties of IDRs[53]. Furthermore, embeddings obtained from diverse PLMs, either individually[54, 55] or in combination[56], demonstrated utility in predicting binding of IDRs to proteins, DNA, RNA, ions and/or ligands[54, 56] [55].

Apart from PLMs, other alignment-free approaches to sequence comparison, including word- and information theory-based metrics, have been investigated[57]. Word-based (also known as k-mer-based) methods are widely used in sequence comparison. However, k-mer based methods face challenges, such as exponentially diminishing sensitivity to remote homologs as the k-mer length increases. To address this issue, the Similarity/Homology Assessment by Relating K-mers (SHARK)[26] sequence comparison algorithm uses an amino-acid distance matrix to quantify the similarity between k-mers and can detect homology for longer k-mers. For pairwise homology assessment, the recently developed SHARK-dive model[26] enables detection of conservation of motifs and regions of different lengths[26]. SHARK-dive was trained on a set of orthologous non-domain regions, where the correspondence between orthologs is determined by the domain architecture of the protein, thereby removing explicit alignment requirements. SHARK-dive outperforms BLAST and HMMER on a systematic homology assessment task, and was shown to be uniquely able to accurately discern homology from several IDR replacement experiments[23, 27, 28]. Thus, SHARK-dive can support homology-based transfer of functional annotations for IDRs. Moreover, we developed SHARK-capture to identify conserved features within sets of IDRs. Once homologous IDRs are identified through SHARK-dive or other methods, SHARK-capture can detect conserved motifs, such as SLIMs that often harbor binding-related functions[58]. As an example, SHARK-capture identified a short (4-residue) functional motif in the C-terminal IDR of the *S. cerevisiae* RNA helicase Ded1p, which, when mutated, reduces ATPase activity by 50%. Another alignment-free approach based on homology is FAIDR, or Functional Annotation of Intrinsically Disordered Regions[28, 59]. FAIDR represents IDRs as vectors of 'evolutionary signatures', which are Z-scores that quantify the degree to which the average value of a sequence feature in evolutionarily related IDRs deviates from a null expectation. The sequence features cover a broad range of compositional and physicochemical properties, as well as repeats, and motifs. Using evolutionary signatures as input, FAIDR annotates a query IDR with specific Gene Ontology (GO) terms and associated likelihood scores. The method is designed to simultaneously optimize the likelihood of functional association at both the protein and IDR levels. Because FAIDR employs regularized logistic regression during training on representations of IDRs as evolutionary signatures, it learns to identify the most predictive and interpretable molecular features associated with specific GO term annotations. FAIDR has been applied proteome-wide on yeast[28] and human IDRs[30], demonstrating predictive potential as validated by independent experimental data. Furthermore, FAIDR showed robust generalization on held-out datasets for a range of functional categories, including DNA-related (e.g., DNA repair, histone binding), RNA-related (e.g., RNA splicing, spliceosomal complex), and membrane-associated GO terms (e.g., G-protein coupled receptor activity, ion transport). For predictions of association with biomolecular condensates, FAIDR achieved precision comparable to that of specialized state-of-the-art methods developed specifically for this task[30]. A key strength of FAIDR is the interpretability of its predictive features, which link specific sequence elements within IDRs to particular molecular functions. The interpretability facilitates the design of targeted validation experiments and can support *de novo* design of IDRs with desired functionalities[60]. Notably, FAIDR's proteome-wide functional annotations indicate that more than half of human IDRs could be associated with three or more functional classes[30], supporting the concept that functional "moonlighting" might be more prevalent in disordered regions than in structured domains[43].

Related and important areas of exploration include the transfer of existing functional annotations between related IDRs, as well as the development of novel representations of IDR sequence and conformational ensemble data that enable their classification. Most recent developments in these directions include improved MSA quality control, and utilization of PLM-based methods to facilitate more accurate transfer of GO annotations between IDRs[61][62]; and leveraging conformational ensembles to capture the dynamic nature of IDRs and correlate conformational properties with function[45, 63]. Additionally, promising efforts have been made to develop representations that capture evolutionary information in related IDRs using self-supervised learning, providing an alternative to MSA-based methods and showing correlations with functional GO annotations[64].

Despite the recent advances of alignment-free approaches, two long-standing questions remain: i) can IDRs be functionally annotated at a finer resolution?, and ii) can functional interrelationships between distinct IDRs, and between IDRs and structured domains, be predicted? Addressing these challenges requires both improved data representations and effective integration of sequence-based, structural, evolutionary, and functional information.

**I See the Light: Illuminating binding sites in disorder**

Over the years, considerable effort has been placed into developing disorder predictors[65] that characterize both the structural[66, 67] and functional properties of IDRs[54, 68]. So far, IDR-specific functional predictions have primarily focused on interaction-related roles (DNA-, RNA-, protein-, ion- and ligand- binding)[54, 69–73] and the six broad functional categories adopted by DisProt8.0 (entropic chains, display sites, chaperones, effectors, assemblers, and scavengers)[74, 75].

SPOT-Disorder2[76], for example, integrates evolutionary information and predicted one-dimensional structural properties of disordered regions to estimate disorder propensities and to identify MORFs within disordered regions. Although the prediction relies on an ambiguous disorder score, this model achieves performance comparable to dedicated MORF predictors, including ANCHOR2[77], DisoRDPbind[78] and MoRFchibi 2.0[79]. Similarly, ANCHOR2 predicts conditional folding of disordered regions upon binding to another protein. It identifies segments that lack sufficient intrachain interactions to fold independently but are likely to become stabilized upon binding to a structured partner. Notably, its performance is largely unaffected by the amino acid composition of the IDR, with all secondary structure propensities are predicted with similar accuracy, regardless of how often they appear in conditionally folding IDRs. The recently published MoRFchibi 2.0 represents the current state-of-the-art in MoRF prediction. When evaluated on an independent dataset of conditionally folding IDRs from curated databases, it outperforms even general protein-binding and MoRF predictors which leverage AlphaFold models and PLMs[79]. To achieve this, MoRFchibi 2.0 combines predictions from multiple convolutional neural networks (CNNs) and applies a statistical correction based on the prevalence of MoRFs in the training data. This adjustment ensures that the output scores reflect meaningful probabilities, which enhances interpretability and enables direct comparison across tools that use the same scoring system.

Several methods focus on prediction of binding sites for nucleic acids or proteins within IDRs. These models explore various machine learning architectures and sequence-based features. For example, DeepDISOBind[70] predicts IDR binding sites for proteins, RNA, and DNA by integrating features such as relative amino acid propensities, secondary structure information from PSIPRED[80], and disorder predictions from SPOT-DisorderSingle[66]. Furthermore, ablation tests have shown that features such as sequence conservation and physicochemical properties significantly improve performance across tasks, e.g., linker prediction by APOD[81], lipid-binding site prediction by DisoLipPred[71], and the prediction of DNA-, RNA-, protein-binding sites, as well as linker identification, by fIDPnn[69]. Additionally, scores for linker propensity and protein/RNA/DNA binding (predicted by earlier methods DFLPred[82] and DisoRDPBind[78]) are incorporated as inputs to DisoLipPred and fIDPnn to further enhance performance. Presently, the only methods that predict lipid-binding residues are DisoLipPred and DisoFLAG[54], and DisoFLAG is also unique in predicting ion-binding residues. Leveraging the strengths of individual methods, DEPICTER2[83] integrates the results of best-performing predictors for each functional category to generate comprehensive predictions across disordered linkers and binding to proteins, peptides, DNA, RNA, and lipids. While the methods discussed in this section provide broad coverage of functionally relevant IDR interactions (e.g., binding to proteins, nucleic acids, or lipids), FAIDR complements these approaches by providing more specific GO-based functional predictions. In the future, integrating both levels of functional annotation could give a more comprehensive understanding and classification of IDR functions across the proteome.

**Be Our Guest: Welcoming new methods in protein function prediction**

Historically, functional annotation of structured proteins and domains has relied on sequence- or structure-based homology inference, using multiple sequence alignments (MSAs), experimental structures, and

protein–protein interaction networks (PPIs) to transfer functional annotations from characterized homologs to uncharacterized proteins. The rapid growth of genomic and AI-predicted structural databases[84] has outpaced functional studies, presenting new opportunities to extract evolutionary information and enable large-scale functional inferences. "The Encyclopedia of Domains" (TED)[38] leveraged AlphaFold2 structures and machine learning tools like Merizo[85] and Chainsaw[86] to annotate over 365 million protein domains. Because structure tends to be more conserved than sequence, TED identified 100 million domains beyond those identified by Pfam. Combined with fast structural comparison tools such as Foldseek[87] and Merizo-search[88], TED now enables large-scale functional prediction and evolutionary analyses, paving the way for structural phylogenetics[89].

The expanding structural and sequence databases provide a rich foundation for current predictive models that integrate both types of information to enhance protein function prediction. For example, recent methods that take structural and sequence data as inputs[90–95] (**Table 1**) typically encode structural information as contact maps, secondary structure annotations, or graphs, while sequences are represented as patterns and long-range dependencies within a single or multiple sequences (**Figure 1**). To learn informative representations, deep learning models commonly employ graph neural networks (GNNs) and CNNs for structural data[90, 92–94] (**Table 1**). Sequences are typically processed using recurrent neural networks (RNNs), long short-term memory networks (LSTMs), transformers, and CNNs, all of which are well-suited for detecting sequence motifs, conserved regions, conservation patterns and features shared across different proteins[90–94, 96–105] (**Table 1**). The architectures are often combined in multi-modal models that integrate not only information from sequence and structure, but also complementary inputs including PPI networks, homology information, textual descriptions of proteins and their functions, multi-omics data etc.[91, 98–110] (**Figure 1**, **Table 1**). While the sequence-based model, DeepGOPlus[111], remains a standard baseline for comparison, multi-modal approaches reportedly provide a more comprehensive view of protein function, often compensating for individual data limitations and improving generalization, albeit with potential noise from separately embedded inputs.

Arguably, the most significant recent advancement in protein function prediction was motivated by the success of large transformer-based language models in the processing of human language (natural language processing, NLP). Applications to protein sequences lead to large protein language models (PLMs)[112–117], which were trained on millions of sequences. Early LSTM-based models already demonstrated implicit learning of structural and functional features, outperforming traditional sequence-based methods in tasks like secondary structure and disorder prediction, and GO annotation transfer[62, 118]. Similarly, PLMs leverage deep learning to encode rich sequence representations that implicitly capture structural and evolutionary information, matching or at times outperforming MSA-based methods in function and contact prediction[113, 114, 116, 119], especially in cases of lower MSA depth[113]. Interestingly, while providing a model with both homology data and PLM embeddings tends to improve predictions across different tasks[93, 96, 97, 99], directly incorporating MSA information during PLM training has reportedly no effect or can even reduce accuracy, especially for prediction of intrinsic disorder[120].

Rich contextual sequence representations generated by PLMs have been shown to improve performance on various downstream tasks, such as fold classification, identification of distant homologs[112, 115], prediction of mutational effects, assignment of GO terms[99, 112–115], per-residue predictions of secondary structure[121], and structure prediction[116]. Consequently, PLM embeddings are now central components of integrative models for protein function prediction and are increasingly combined with diverse auxiliary data sources such as sequence homology data[103][96, 104], secondary structure[105], residue contact maps [90, 91, 94], structure graphs[92, 93], template modeling (TM) scores[108] and PPI networks[101, 102] (**Table 1**). While no single PLM architecture has proven universally superior across functional prediction tasks[122], larger transformer-based models have demonstrated the best generalization to rare proteins and underrepresented functional classes. This could be because different relevant features get encoded at different depths (i.e., layers) of the transformer model. For instance, while binding sites and PTMs have been shown to be represented across multiple attention layers, structural features are predominantly encoded in deeper layers[123]. This layered information hierarchy suggests that models with more layers can capture progressively more nuanced aspects of protein structure and function. This might explain the trend toward increasingly deep and complex architectures for PLMs and protein function prediction models in general, as reflected by the examples listed in **Table1**.

Notably, recent studies have highlighted a significant bias in protein sequence datasets, which are heavily skewed toward certain species[124]. This bias likely propagates into PLM embeddings, limiting their generalizability, especially for rare or underrepresented proteins. At the same time, no single model outperforms others across tasks, and annual improvements among newly published models have become relatively marginal. This might suggest that current approaches are reaching a performance ceiling given the available input data. To test this hypothesis, universal assessments and standardized benchmarking efforts are essential[122, 125–127] (**Box 2**). Based on current trends, future improvements may depend less on increasing model size and more on improving the diversity and quality of training data and evaluation benchmarks[114, 124, 125, 128]. Future progress could also come from optimizing input data representations in integrative models to reduce biases introduced by separately embedding each data type, as well as from rethinking the formulation of protein function prediction.

**Out There: Limitations of state-of-the-art models on IDRs**

The models discussed above have collectively led to significant improvements in functional prediction for structured proteins. However, their performance on IDPs and IDRs remains largely unassessed. When evaluating predictions from four state-of-the-art protein function predictors with different model architectures and supported data input types (DeepGOPlus[111], DeepFRI[90], StarFunc[95] and Sprof-GO[104]), we observed that the overall information content[129] of the predictions decreased as disorder content increased (**Figure 2**). Most strikingly, for IDPs that by definition lack any stable structure, all predictors returned few confident predictions. Those that were returned typically contained GO terms with little to no information content. This is likely due to multiple factors. First, functional prediction models are rarely trained or evaluated on IDRs. Second, many of the models might rely heavily on structural data, which becomes ineffective when structural plasticity is high and stable tertiary and secondary structure is absent, as is inherent to IDRs. Similarly, reliance on sequence homology likely hinders performance on more divergent IDRs/IDPs, for which MSAs are less informative.

In contrast, the methods that focus on predicting broad IDR functional categories, binding sites, conformational properties, and IDR-specific GO-terms (FAIDR) are trained on features of disordered regions. When contrasted against the aforementioned state-of-the-art functional predictors, FAIDR yields markedly more informative predictions on IDRs and IDPs (**Figure 2**). This indicates that IDR-containing proteins, and IDPs, remain a challenge for general state-of-the-art protein function prediction methods, even in cases where a conditional fold is known to occur in an IDR (**Figure 2**).

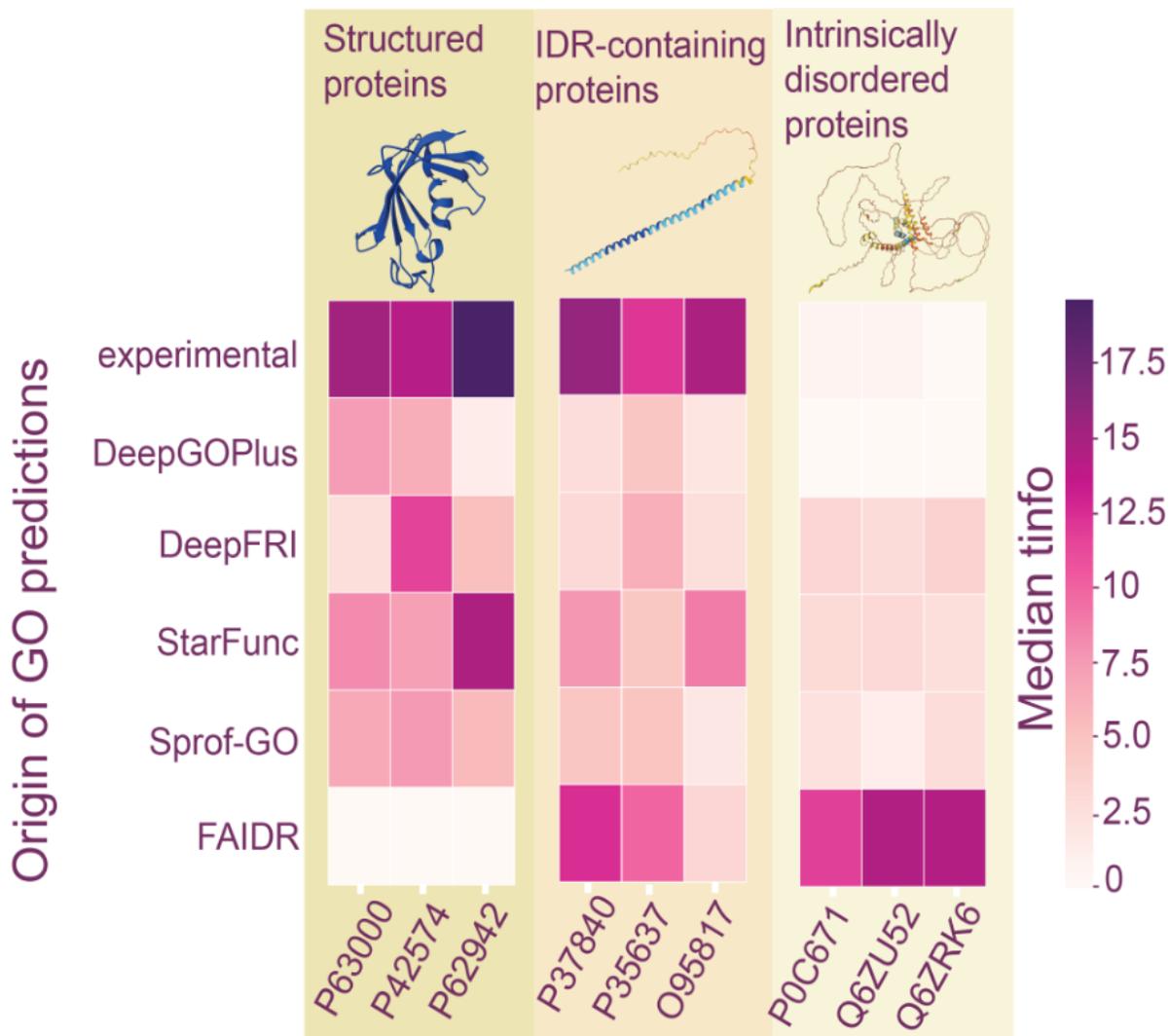

**Figure 2. Comparison of protein function prediction models across structured and disordered proteins.** This figure represents a heatmap of cumulative median information content (*tinfo*)[129] of Gene Ontology (GO) predictions for all three aspects (molecular function, biological process and cellular component). The color scale corresponds to the value of *tinfo*, which quantifies the depth of functional annotation within the GO hierarchy. Higher *tinfo* (e.g. 12) reflects more specific and complete functional annotation, while lower *tinfo* (e.g. 5) describes broader GO terms, usually higher in the GO tree. The heatmap includes six rows: experimentally determined GO annotations, and protein function predictions of four state-of-the-art models[90, 95, 104, 111] and of FAIDR[28, 30], a model trained to annotate intrinsically disordered regions and proteins (IDRs and IDPs). The columns represent proteins grouped into fully structured (left), IDR-containing proteins or conditionally folding IDR (middle), and fully disordered proteins (i.e., IDPs) (right). The Uniprot IDs of tested proteins are listed at the bottom of the figure. Representative structures, obtained from AlphaFold database, for each protein category are shown at the top of the figure (left to right: P62942, P37840, P0C671). Alpha-synuclein is an IDP that conditionally folds upon binding to lipid membranes and is included here because of its deeper MSA[50]. State-of-the-art models perform well

for fully structured proteins but struggle with IDR-containing proteins and IDPs. FAIDR, trained specifically on IDRs and IDPs, offers the most specific and informative functional predictions for IDPs. The strong performance of current models on structured proteins, but their difficulty in handling proteins with disorder, underscores the need for integrative models capable of making functional annotations across all protein types.

While disorder-aware approaches appear to capture properties of IDRs that correlate with their functions, the overall functional categorization of IDRs still remains more limited in scope and detail compared to that of structured proteins. This discrepancy reflects the lack of high-throughput functional annotations for IDRs (*vide infra*) and the challenges with sequence-based annotation transfers for IDRs. Indeed, DisProt and MobiDB community curators have identified disorder function annotation as a bottleneck in IDR research, and efforts are already underway to expand the currently available annotations[130, 131].

Beyond functional prediction, our understanding of the IDR sequence-ensemble-function relationship has recently been significantly advanced by other dedicated ML-based methods[45, 63, 132–135]. These advances have enabled the *de novo* design of IDR sequences with tailored conformational properties[136], as well as the generation of IDR ensembles that are both geometrically accurate[135, 137] and restrained by experimental data[138–140] (reviewed comprehensively in[141]).

**Part of That World: Building a framework for IDR functional prediction and validation**

For over 30 years, organized competitions have catalyzed innovation in several aspects of protein structure and function prediction and gained wide adoption by both computational and experimental communities (**Box 2**). The transformative impact of these challenges stems from their ability to engage the community through both competitive and collaborative spirit. Their added value comes in the creation of shared databases, benchmarks, standardized experimental datasets and uniform evaluation criteria for the computational methods. The resulting structured (big) data and well-defined objectives also provide an ideal basis for current trends towards AI-based methods development.

The Critical Assessment of Structure Prediction (CASP)[142] was the first community challenge aimed at advancing macromolecular structure prediction from sequence information. This annual competition led to breakthroughs in structural prediction at atomic-resolution accuracy for single proteins and domains, driven by deep learning methods, particularly AlphaFold2[84]. Nevertheless, IDRs are not included in the evaluation tasks, as current tools do not consider structural ensemble prediction. This is primarily due to limited availability of experimental data, which are generally sparse and mostly restricted to specialized NMR studies[3]. Similarly, IDRs are largely left out of the Critical Assessment of PRediction of Interactions (CAPRI) challenge[143], which has been running since 2000 and evaluates the accuracy and effectiveness of protein docking algorithms, which aim to predict the three-dimensional structures of protein complexes from the structures of their individual components.

On the other hand, the precise identification of IDRs is a well-defined task that is assessed by the Critical Assessment of Intrinsic protein Disorder (CAID) (**Box 2**)[144][145], which compares sequence-based disorder predictors since 2018. However, identifying structural or functional properties beyond a binary classification of disordered or ordered is not yet a well-defined challenge and may require concerted efforts before it could be formalized and standardized.

The Critical Assessment of protein Function Annotation algorithms (CAFA) (**Box 2**)[146], running since 2010, has facilitated a large-scale unbiased assessment of computational methods dedicated to predicting protein function. Protein function is associated with GO and HPO (Human Phenotype Ontology) terms in three categories: molecular function, biological process, and cellular component. Thus far, DisProt, which has pioneered ontologies for IDR function (**Box 2**), has only ~3000 entries and three main functional categories (entropic chain, molecular recognition display site, self-regulatory activity)[130]. These annotations are also integrated into MobiDB which now provides predicted functional annotations[131]. We advocate for a community effort to formalize a "dis-CAFA" challenge that stimulates the evaluation and development of algorithms that are designed for the prediction of IDR functions. While computational predictions of IDR functions already exist, as we have reviewed above, their critical assessment and broad-scale testing lags behind. To further expand these efforts, we propose the development of a disordered function ontology and large-scale experimental annotation of disordered functions. Starting efforts in this direction have been undertaken by the DisProt and MobiDB communities, as well as the COST initiative (European Cooperation in Science and Technology) ML4NGP[131]. Similarly, crowd-sourcing data collection initiatives are advancing the annotation and prediction of biomolecular condensates [147]. In this manner, a community-led effort to further map sequence-to-function relationships in IDRs is expected to drive the next generation of technologies and stimulate new biological insights.

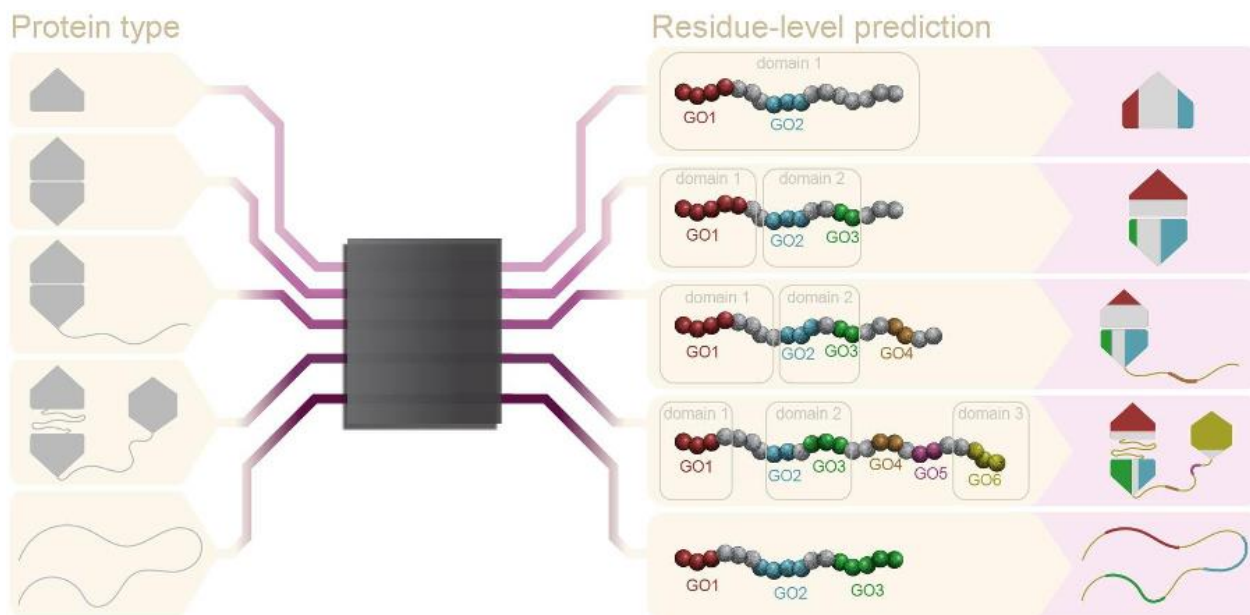

**Figure 3. Future directions for protein function prediction models.** This figure illustrates the envisioned framework for next-generation protein function prediction models. On the left, different structural classes of proteins are shown, ranging from fully ordered single- and multi-domain proteins to those that contain intrinsically disordered regions (IDRs) or are entirely disordered (IDPs). The model processes inputs from proteins with varying degrees of disorder and provides residue-level functional annotations across all structural cases. On the right, predicted functions are visualized, with different colors that highlight distinct Gene Ontology (GO) annotations (GO1 to GO6) mapped onto both sequence and structural representations. This approach would enable a more precise and comprehensive understanding of molecular function beyond current whole-protein predictions.

**Conclusions and Outlook**

IDRs lack stable secondary and tertiary structures, but play essential roles in cellular regulation, signaling, and disease. However, IDPs and IDRs still pose challenges for functional prediction due to their rapid sequence evolution, limited positional conservation, and complex sequence-ensemble-function relationships. Due to these challenges, current methods and trends in protein function prediction largely dismiss intrinsic disorder. Recent methodological advances—spanning protein language models (PLMs), alignment-free methods like SHARK-dive and SHARK-capture, and IDR-specific tools such as FAIDR— have begun to bridge the gap. These tools capture both local motifs and global sequence features and have demonstrated potential in providing more informative annotations for IDRs than general-purpose predictors. Nevertheless, most state-of-the-art models remain biased toward structured regions and underperform on disordered proteins.

The functional annotation of IDRs presents challenges stemming from multiple compounding factors: annotation inequality, limited availability of high-throughput IDR-focused studies, barriers in data curation and standardization, and the inherent complexity of IDR structural ensembles and functional diversity. Annotation inequality arises from a historical focus on a small subset of well-characterized proteins—often disease-associated—resulting in uneven functional coverage. For instance, an estimated 75% (95%) of all papers on protein-coding genes have concentrated on fewer than 10% (25%) of human proteins, with nearly half of human genes effectively absent from the literature[148, 149]. This bias leaves many proteins, particularly those containing IDRs, underrepresented in the scientific literature. Furthermore, dedicated high-throughput studies that focus on IDRs specifically are rare, albeit recent trends point in a promising direction[150]. Indeed, recent high-throughput assays performed on different types of IDRs, including activation domains and mRNA-binding regions, have uncovered new insights into sequence-function relationships in IDRs[151–153].

Further complicating prediction efforts, IDRs often contain overlapping functional elements, and their behavior is highly context-dependent—modulated by binding partners, cellular states, alternative splicing, and post-translational modifications. These factors create a complex functional landscape that defies simple classification. Mutations in IDRs are frequently implicated in cancer and neurodevelopmental disorders [4], underscoring the clinical importance of developing robust tools for their annotation and analysis. Functional prediction methods that address these challenges could pave the way for more informed protein design and therapeutic discovery (**Figure 3**). The underrepresentation of IDRs in training data, benchmarks, and community challenges (e.g., CASP, CAFA, CAPRI) continues to limit progress. This underscores the urgent need for disorder-aware evaluation frameworks, standardized ontologies, and high-throughput functional annotation efforts. A dedicated challenge for IDRs—"dis-CAFA"—could catalyze this shift by promoting method development, data curation, and evaluation tailored to the biology of disorder.

Looking ahead, future improvements in functional prediction will likely depend less on increasing model size and more on diversifying training data, improving integrative representations, and refining problem formulations that respect the unique biophysical and evolutionary properties of IDRs (**Outstanding Questions**). As illustrated in **Figure 3**, future models that can provide high-resolution functional annotations across the full continuum of protein conformational space, i.e., from fully ordered to fully disordered states, will significantly aid in this endeavor. Such models would move beyond current protein-level annotations towards mapping specific Gene Ontology (GO), or other functional terms (vide infra), directly onto defined residue stretches of a protein sequence. Such region-level mapping would deliver a more precise and comprehensive understanding of protein function. For example, enabling multi-label classification, considering the GO hierarchy in prediction outputs, aiming for finer-grade level annotation and annotating rare GO terms all represent potential future avenues. Unlocking the full functional potential of disordered regions will not only complete the sequence-to-function map but also open new avenues in cellular biology, protein engineering, and drug discovery.

**Table 1. Selected top performing methods in CAFA and their input data types and accessibility.** The Notes column includes information about the type of language model or embeddings that are used.

| Main input type | Model name (year) | Input data type | | | | | Server/ Code | Notes | ref |
|---|---|---|---|---|---|---|---|---|---|
| | | Sequence | Sequence features, domain info | Structure | PPI | Other input type | | | |
| Sequence | **ATGO** (2022) | X | | | | | Server | Large PLM[a] embeddings | [96] |
| | **TEMPROT** (2023) | X | | | | | GitHub | Large PLM embeddings | [97] |
| Sequence and structure | **deepFRI** (2021) | X | | X | | | Server | / | [90] |
| | **LM-GVP** (2022) | X | | X | | | GitHub | Large PLM embeddings | [92] |
| | **TransFun** (2023) | X | | X | | | GitHub | Large PLM embeddings | [93] |
| | **Struct2GO** (2023) | X | | X | | | Server | Small PLM embeddings | [94] |
| Sequence-based integrative models | **Graph2GO** (2020) | X | X | | X | X (subcellular localization) | GitHub | / | [98] |
| | **TALE** (2021) | X | | | | X (GO labels only for training) | GitHub | / | [99] |
| | **NetGO2** (2021) | X | X | | X | X (textual description of proteins) | Inactive server | / | [100] |
| | **GAT-GO** (2022) | X | X | X | | | / | Large PLM embeddings | [91] |
| | **ProTranslator** (2022) | X | | | X | X (protein and GO functions description during training) | GitHub | Biomedical LM[b] (PubMedBert[154]) | [101] |
| | **PANDA2** (2022) | X | | | | X (GO info and sequence similarity) | Server | Large PLM embeddings | [103] |
| | **NetGO3** (2023) | X | X | | X | X (textual description of features) | Server | Large PLM embeddings | [102] |
| | **SPROF-GO** (2023) | X | | | X | X (homology info) | Server | Large PLM embeddings | [104] |
| | **DeepSS2GO** (2024) | X | | X | | X (homology info) | GitHub | Large PLM embeddings | [105] |
| | **StarFunc** (2024) | X | X | X | X | X (homology info) | Server | Large PLM embeddings | [95] |

| | | | | | | | | |
|---|---|---|---|---|---|---|---|---|
| | **DeepGraphGO** (2021) | x* | x | | x | **x** (homology info) | GitHub | / | [106] |
| | **DeepGOZero** (2022) | x | x | | | **x** (GO axioms during training) | GitHub | / | [107] |
| | **ProteinVec** (2023) | x* | | x | | **x** (gene and domain info) | GitHub | Large PLM embeddings | [108] |
| | **LATTE2GO** (2023) | x* | x | | | **x** (multi-omics data, GO data during training) | / | / | [109] |
| | **Domain-PFP** (2023) | x | x | | | **x** (GO labels during training) | Google Colab | / | [110] |
| Other integrative models | **MELISSA** (2024) | | | | x | **x** (GO labels during training) | GitHub | / | [155] |

[a] PLM - protein language model
[b] LM - language model
* implicit use of sequence through extraction of domains or other sequence features

**Glossary**

**Intrinsically disordered region (IDR) and protein (IDP)** - polypeptide segments or entire proteins that do not adopt a stable secondary or tertiary structure under native conditions, instead populating dynamic ensembles of rapidly interconverting conformations.

**Convolutional neural networks (CNNs)** - a class of neural networks originally developed for the analysis of grid-like data, such as images and videos. These neural networks use convolutional layers to detect local patterns and are commonly employed in deep learning. CNNs have also been successfully applied to protein sequences and structures to identify patterns and extract features for predicting protein function and structure.

**Deep learning (DL)** - a subset of machine learning that uses multi-layered neural networks to automatically learn representations from large volumes of data. Different neural network architectures (e.g., CNNs, RNNs, LSTMs, and Transformers) have been developed for different types of data.

**Graph neural networks (GNNs)** are neural network models designed to operate on graph-structured data, capturing relationships between nodes and edges. In biology, GNNs are commonly used to encode representations of protein structures, molecular interaction networks, and hierarchical ontologies, such as Gene Ontology (GO) graphs.

**Recurrent neural networks (RNNs)** - a class of neural networks designed to process sequential data by maintaining a memory of previous inputs. This allows RNNs to model context-dependence, e.g. by predicting the next amino acid in a protein sequence based on the preceding ones.

**Long short-term memory networks (LSTMs)** - a specialized type of RNN designed to capture and retain long-term dependencies in sequential data (e.g. protein sequences), by addressing issues such as the vanishing gradient problem.

**Transformers** - neural network architectures that use attention mechanisms to weigh the importance of residues and capture both local and global patterns in sequences.

**Protein language models (PLMs)** - transformer-based deep learning models trained on a large number of diverse protein sequence datasets to learn patterns related to structure, evolution, and function. The resulting fixed-length embeddings are numerical representations that capture rich contextual properties of each protein sequence.

**Gene ontology (GO)** - a structured vocabulary for describing protein functions, organized into three categories: molecular function (MF), biological process (BP), and cellular component (CC). The terms are connected by hierarchical relationships that form a directed graph, in which each term can have multiple parent terms.

**Information content (tinfo)** - an estimated measure of the specificity of a GO term. It is calculated as the negative logarithm of the term's frequency of annotations, or it is inferred from the number of descendants of the term in the GO hierarchy. Terms with higher information content (i.e., fewer descendants) are considered more specific, i.e. of higher information content.

**Multiple sequence alignment (MSA)** - alignment of three or more protein (or nucleotide) sequences to reveal positional conservation, from which evolutionary constraints, structural features and functional importance can be inferred.

**Short linear motifs (SLiMs)** - short, conserved stretches of 3-10 residues within IDRs that can mediate protein-protein interactions, signal transduction, and regulatory functions.

**Molecular recognition motifs (MoRFs)** - short sequence motifs within IDRs that undergo disorder-to-order transitions upon binding to interaction partners, typically other proteins.

**Protein–protein interaction (PPI) networks** - graphs that represent the physical or functional associations between proteins within a cell or organelle. In these graphs, nodes correspond to proteins and edges denote interactions, which are derived from experimental characterization or computational prediction.

**Template modelling (TM) score** - a measure of structural similarity between two protein structures, which is length-independent and generally more robust than root-mean-square deviation (RMSD). TM-scores range from 0 to 1, with higher scores indicating greater structural similarity.

**Text Boxes**

**Box 1. Local and global sequence properties of IDRs encode specific and bulk functions (399 words)**

**Local sequence properties confer functional specificity**
Short linear motifs (SLiMs), also known as eukaryotic linear motifs (ELMs), are short sequences (typically of 3–10 amino acids) found within IDRs. A well-known example is the nuclear localization signal (NLS). Some SLiMs may fold into stable structures upon binding to an interaction partner (MoRFs). SLiM can act as binding sites for other proteins, targets for PTMs, or signals to control or regulate subcellular localization and protein degradation. While SLiMs are generally more conserved than their flanking sequence regions, they can tolerate conservative substitutions that preserve their physicochemical properties. SLiMs are often represented as sequence logos or regular expressions. The overall frequency of a SLiM within an IDR can be conserved rather than the actual SLiM position.

**Global or bulk/low-resolution properties confer global functions**
Global sequence properties, such as biased amino-acid composition, overall charge, or patterning of charged residues can play a critical role in determining the structural and functional characteristics of an IDR. These properties influence global features, including chain compactness (e.g., radius of gyration), conformational entropy, and hydration patterns. The bulk properties have been shown to support IDR functions such as acting as "entropic chains", regulating protein stability, or protecting against environmental stresses like desiccation and freezing.
Global sequence properties also mediate interactions, particularly multivalent binding, which is essential for processes such as biomolecular condensate formation. These interactions are often driven by distributed sequence patterns, rather than precise sequence order, resulting in chemical specificity. Consequently, many sequence variants can encode the same biophysical behavior, leading to degeneracy. Traditional sequence alignment methods often fail to detect these global, functionally relevant properties, highlighting the need for alternative approaches to study them.

**Overlap and interchangeability of global and local function**

Some IDRs have been shown to function through one of the above mechanisms, i.e. either globally (e.g. the shuffling of amino acids retains the overall function) or locally (IDRs only require a SLIM that acts in an autonomous fashion; as shown by SLiM implantation experiments where function is retained). However, most IDRs likely harbor a complex mix of global and local contributions to their functions. It was recently demonstrated that even the same binding function can be achieved through either local or global sequence properties[156].

**Box 2. Resources for protein function prediction**

**Gene ontology (GO):**

GO is a resource containing descriptors of gene functions, cellular locations, and biological processes across different species. GO uses a standardized vocabulary and structured hierarchy to label and organize gene annotations across "molecular function", "subcellular localization", and "biological pathway". GO terms are commonly used for cross-species comparisons, transfer of annotations and functional predictions.

**DisProt database:**

DisProt is a manually curated database that contains information about the sequences, conformational ensembles, and functions of IDRs and IDPs. The database provides disorder annotations that are supported by different experimental sources. It defines IDRs as segments of at least ten consecutive residues that are likely to be linked to specific biological functions and excludes short loops. DisProt also annotates disordered interaction interfaces, including protein-protein and nucleic acid-protein interactions, and annotates linker regions.

**DisProt ontology:**

DisProt Ontology is designed to annotate IDPs and IDRs using a combination of three distinct ontologies: the Intrinsically Disordered Proteins Ontology (IDPO), Gene Ontology (GO), and Evidence and Conclusion Ontology (ECO). The IDPO describes conformational states, structural transitions (e.g. disorder-to-order), and broad disorder-related functions of IDPs/IDRs (e.g. "entropic chain", molecular recognition, self-activation etc.). GO terms are used to describe more specific functional aspects of disordered regions. ECO provides information on the evidence and techniques used to support the annotations.

**CAFA:**

The Critical Assessment of protein Function Annotation algorithms (CAFA) is a community-wide challenge that evaluates computational methods for protein function prediction. In the first part of the challenge, participants receive a set of protein sequences with no known experimental functional annotation and have to submit their GO-term predictions. Over the following months, new experimental annotations for some of these proteins become available and are used as a benchmark to evaluate the accuracy of predictions made by competing models. Since its first edition in 2010, five CAFA challenges have been held, with the most recent one (CAFA 5) in 2022.

**CAID:**

The Critical Assessment of Protein Intrinsic Disorder Prediction (CAID) is a community-driven benchmarking experiment that was established in 2018 to evaluate the emerging models for prediction of IDRs and disordered residues involved in binding. Using the DisProt database as the ground truth, CAID first standardizes the methods and then assesses them based on their accuracy and computational efficiency.

**Acknowledgements**

A.T.P acknowledged funding by the Max Planck Gesellschaft (MPG) "free-floater" RGL funds and by the European Research Council (CONDEVO, ERC grant agreement number 101116284). T.R.A. acknowledges funding from the Initiative and Networking Fund of the Helmholtz Association (Helmholtz Investigator Grant VH-NG-20-14).